\documentclass[aps,prl,twocolumn,groupedaddress,nofootinbib,showpacs]{revtex4}
\usepackage[dvips]{graphicx}

\usepackage{graphicx,epsf,amssymb,amsbsy,amsfonts,amssymb,amsmath}
\usepackage{amssymb}
\usepackage{bbm}

\usepackage[usenames, dvipsnames]{color}

\newif\ifdraft
\drafttrue
\newif\ifpreprint

\preprintfalse

\def \a {\alpha}

\def\eqn#1{eq.~(\ref{#1})}

\def\spa#1.#2{\left\langle#1\,#2\right\rangle}
\def\spb#1.#2{\left[#1\,#2\right]}
\def\spash#1.#2{\spa{\smash{#1}}.{\smash{#2}}}
\def\spbsh#1.#2{\spb{\smash{#1}}.{\smash{#2}}}
\def\sand#1.#2.#3{%
\left\langle\smash{#1}{\vphantom1}^{-}\right|{#2}%
\left|\smash{#3}{\vphantom1}^{-}\right\rangle}
\def\sandpp#1.#2.#3{%
\left\langle\smash{#1}{\vphantom1}^{+}\right|{#2}%
\left|\smash{#3}{\vphantom1}^{+}\right\rangle}
\def\sandpm#1.#2.#3{%
\left\langle\smash{#1}{\vphantom1}^{+}\right|{#2}%
\left|\smash{#3}{\vphantom1}^{-}\right\rangle}
\def\sandmp#1.#2.#3{%
\left\langle\smash{#1}{\vphantom1}^{-}\right|{#2}%
\left|\smash{#3}{\vphantom1}^{+}\right\rangle}

\def\be{   \begin{equation} }
\def\ee{ \end{equation}  }
\def\bea{\begin{eqnarray}}
\def\eea{\end{eqnarray}}
\def\ba{\begin{eqnarray}}
\def\ea{\end{eqnarray}}

\def \a {\alpha}

\def\s{\sigma}

\def\a{\alpha}

\def\q{\theta}

\def\s{\sigma}

\newcommand{\ad}{{\dot{\alpha}}}                           
  
\newcommand {\cW}{{\cal W}}

\newcommand {\cZ}{{\cal Z}}

\newcommand {\cMbar}{ {\overline {\cal M} }}

\newcommand {\cWbar}{ {\overline {\cal W} }}

\newcommand {\SD}{ { {\cal D}^4 } }

\newcommand {\SDbar}{ { {\overline {\cal D}}^4 } }

\newbox\charbox
\newbox\slabox
\def\s#1{{      
        \setbox\charbox=\hbox{$#1$}
        \setbox\slabox=\hbox{$/$}
        \dimen\charbox=\ht\slabox
        \advance\dimen\charbox by -\dp\slabox
        \advance\dimen\charbox by -\ht\charbox
        \advance\dimen\charbox by \dp\charbox
        \divide\dimen\charbox by 2
        \raise-\dimen\charbox\hbox to \wd\charbox{\hss/\hss}
        \llap{$#1$} }}

\begin{document}
\hfuzz 20pt

\noindent SU-ITP-13/03

\vskip-0.8cm
\title{ Hidden Supersymmetry May Imply Duality Invariance}
\vskip0.8cm
\author{ John Joseph M. Carrasco and Renata Kallosh}

\affiliation{
${}$\hbox{Stanford Institute for Theoretical Physics and  Department of Physics,}\\ 
\hbox{Stanford University, Stanford, CA 94305, USA}
}


\begin{abstract}
We provide evidence that a particular hidden supersymmetry, when combined with half-maximal deformed global supersymmetry,  implies that the theory is invariant under duality rotations of the vector and spinor fields.  Based on a complete 8+8 supersymmetric model constructed recently, we argue that this hidden supersymmetry happens if and only if there is a Born-Infeld dependence on the Maxwell field strength and  a Volkov-Akulov dependence on the Goldstino, up to local non-linear field redefinitions.  We have tested our proposal for  the $\mathcal{N}=2$ superfield action with manifest $\mathcal{N}=2$ supersymmetry and  hidden $\mathcal{N}=2$ supersymmetry at the level ${\cal W}^{10}$, the highest level of deformation known for this model. We have established that it is  $\mathcal{N}=2$ self-dual, although the self-duality was not  required in the original construction of this model.  Highlighting the utility of considering duality-conserving sources of deformation, we can verify this invariance directly in an alternate construction of this very same action. 
\end{abstract}

\pacs{11.15.Bt, 11.30.Pb, 11.55.Bq \hspace{1cm}}

\maketitle



The idea that electric and magnetic field densities can be placed on equal footing such that theories may be invariant when these fields are rotated amongst each other is called {\it duality invariance}.  Obvious in sourceless equations of motion, this is a very physical symmetry to make manifest in gauge theories. While easy to break, this symmetry lends itself beautifully  to the emergence of Poincare invariance~\cite{Bunster:2012hm}.  This natural idea and generalizations have been the cause of many intriguing studies since the dawn of modern electromagnetic theory perhaps most famously with the Born-Infeld (BI) theory~\cite{BI,Schrodinger,Dirac:1962iy,Volkov:1973ix,Deser:1976iy,Gaillard:1981rj,Cecotti:1986gb,Henneaux:1988gg,Gibbons:1995cv,Bagger:1996wp,Gaillard:1997rt,Ketov:1998ku,Rocek:1997hi,Tseytlin:1999dj,Kuzenko:2000tg,Kuzenko:2000uh,Bellucci:2001hd,Ivanov:2002ab}.  It was in fact Schr\"odinger \cite{Schrodinger} himself who first pointed out that the BI theory has a duality symmetry.
Recent excitement has arisen again~ \cite{Bossard:2010bd,Beisert:2010jx,Kallosh:2011dp,BN,CKR,CKO,Neq2duality,Pasti:2012wv,Roiban:2012gi,Kuzenko:2012ht,Bellucci:2012nz,Ivanov:2012bq,Kuzenko:2013gr,Aschieri:2013nda},  due in part to arguments suggesting that duality symmetry may play an important role in explaining the ultraviolet behavior discovered in some of the recent loop computations in supergravity~\cite{threeLoop,threeLoopManifest,fourLoop,fourLoopManifest,NeqEightBehavior,threeLoopNeq4,Kallosh:2012ei,Bossard:2012xs}.    Here we disentangle mysteries connecting supersymmetric abelian gauge theories\footnote{This is not as unrelated as one might imagine.  Recent arguments~\cite{bcj, bcjLoop, bcjUV,n4U1anom}, have suggested that the path forward to understanding gravity theories may lie in thoroughly understanding gauge theories.}, hidden symmetries, and $U(1)$ duality invariance.  

There is value to considering various ways of encoding symmetry-respecting deformations of theories to allow additional interaction terms, or outright counterterms.  A  powerful and familiar tool is the introduction of auxiliary fields~\cite{Ivanov:2012bq,Kuzenko:2013gr}.  Here we apply an alternative very compact algorithmic notion introduced in ref.~\cite{BN}, and generalized in refs.~\cite{CKR,CKO,Neq2duality}, namely the specification of a source of deformation at a particular order, whose ramifications on the action at all orders can be recursively determined.  The deformation sources themselves can be used to distinguish theories that share certain symmetries but cannot be related through local field redefinitions.  This represents an extraordinary compression of the action.  Our proof here of the duality invariance of a particular ${\cal N}=2$ theory will arise constructively by specifying exactly such deformations relevant to build the model to the known order.    We will explain that this self-duality is implied by the hidden $\mathcal{N}=2$ supersymmetry of this model. 

Much inspiration comes from the fact that there are infinitely many distinguishable  self-dual actions\footnote{Examples of the complicated relations between various versions of  ${\cal N}=1$ Goldstino models are given  by  comparisons of the  ${\cal N}=1$ Volkov-Akulov models with related versions including the Rocek model, Casalbuoni-De~Curtis-Dominici-Feruglio-Gatto model, the supersymmetric Born-Infeld model, and a more recent Komargodski-Seiberg model (see~\cite{Kuzenko:2011tj} and refs.~therein).}  with  manifest  ${\cal N}=1$ supersymmetry, however, only the ones related to the Volkov-Akulov fermionic action by a local field redefinition have an additional hidden supersymmetry~\cite{Kuzenko:2005wh}.     Most notably, ref.~\cite{Kuzenko:2005wh} demonstrated that the fermionic component of the duality-satisfying ${\cal N}=1$ BI model can be locally deformed into the Volkov-Akulov action.     We will argue that this structure generalizes to the ${\cal N}=2$ supersymmetric case.   Namely we provide evidence that from a large class of the recently constructed manifestly  $\mathcal{N}=2$ supersymmetric $U(1)$ self-dual models \cite{Neq2duality} only the one with Born-Infeld dependence on the Maxwell field strength and  Volkov-Akulov type dependence on the Goldstino is consistent with this particular extra hidden supersymmetry,  up to field redefinitions.    Our general argument, based on the action given in  \cite{BCKSV}, will be supported by an example of the manifestly $\mathcal{N}=2$ supersymmetric model with $\mathcal{N}=2$ hidden supersymmetry \cite{Bellucci:2001hd} where the action at the superfield level ${\cal W}^{(10)}$ is available\footnote{The ${\cal W}^{10}$ level was given implicitly in~\cite{Bellucci:2001hd}, and we thank S. Belluci for providing also an explicit expression in recent communication.}.   Prior to the calculation presented here it was not known if it is also self-dual (although it was conjectured in ref.~\cite{Bellucci:2001hd} to be so).

To appreciate the use of duality consistent sources of deformation, it is helpful to begin by considering the purely bosonic case.  One way~\cite{CKR}  to find  ${\mathcal N}=0$ duality satisfying theories is to apply the Noether-Gaillard-Zumino (NGZ) condition \cite{Gaillard:1981rj,Gaillard:1997rt}
\be
F\tilde F+ G\tilde G=0
\label{GZ}\ee
in the form of the Courant-Hilbert equation
\be
{\cal L}_x^2- {\cal L}_y^2=1
\label{CH}\ee 
where $\tilde G^{\mu \nu}=2  \frac{ \partial {\cal L}(F)}{\partial F_{\mu \nu}}\, , 
t= {\frac{1}{4}} {F^2}\, , z= {\frac{1}{4}} F\tilde F\, , 
x=t\, ,   y= \sqrt {t^2+ z^2} 
$.
Requiring analyticity of ${\cal L}$ for small values of $F$ -- one may use an ansatz \cite{CKR} 
\be
\label{lagrangeAnsatz}
{\cal L}=\Big(g^{-2} \!\!\!\!\!  \sum_{m=0, p=0} \!\!\!\!\!  
  g^{2 (p + 2 m)} c_{(p, 2 m)} t^p z^{2 m}\Big)- c_{(0,0)} g^{-2} \, ,
\ee
and solve \eqn{CH} algebraically, order by order in $g^2$, fixing the constant coefficients $c_{(i,j)}$.  One set of solutions of the self-duality condition is the Born-Infeld Lagrangian
 \be
 \label{BILag}
 {\cal L}_{\rm BI}=g^{-2}(1 - \sqrt{\Delta}) 
\ee
where $g$ is the coupling constant,  and $\Delta = 1 + 2 g^2 (F^2/4) - g^4 (F \tilde{F}/4)^2$.   
Using $t, z$   variables, one can rewrite the Born-Infeld Lagrangian simply as
\be
{\cal L}_{\rm BI}=g^{-2}(1 - \sqrt{1 + 2 g^2 t - g^4 z^2\,})\, .
\ee

This procedure can be understood as introducing a generic sum over sources of duality-satisfying deformations~\cite{CKR}
\be{\cal I}(T^{-},  T^{*+})=  \sum_{n=0} ^{\infty} {a_n\over 8 g^2} \Big( {\textstyle \frac{1}{4}}\, g^4  ({ T^*}{}^{+})^2(T^-)^2 \Big)^{n+1} \,  
\ee
 where $T=F -i G$ and $ T^*=  F+i G$. It
 leads to a twisted non-linear self-duality condition
\be\label{biAnsatz}
T^+_{\mu \nu} =  {g^2\over 16}  {T^*}{}^{+}_{\mu\nu} (T^-)^2  \Big[\, 1 + \sum_{n=0} d_n \Big( {\textstyle \frac{1}{4}}\, g^4  ({ T^*}{}^{+})^2(T^-)^2 \Big)^{n} \,  \Big],
\ee
Here $a_0 = 1+d_0$ and $a_n= {d_n\over n+1}$ for $n\geq 1$, where $d_{n}$ are arbitrary  real parameters. For BI the sum is a hypergeometric function\footnote{Recently understood in ~\cite{Aschieri:2013nda} as a function satisfying a hidden quartic equation of the Schr\"odinger construction of the BI model.}~\cite{CKR}.  Another model, introduced by Bossard-Nicolai~\cite{BN,CKR}, has all vanishing $d_n$.  No local change of field variables $A_\mu' = A_\mu' (A_\mu)$ relates these theories.

The perturbative relation between $G$ and $F$ follows from \eqn{biAnsatz} and 
an action depending only on $F$, where $G$ is the functional of $F$, has the reconstructive form~\cite{CKO}
\be
S(F)= {1\over 4 g^2} \int d^4 x\, d\left(g^2\right) \, F \tilde G\,.
\ee
For any choice of $d_n$ the action has the self-duality property. 

 It should be noted that the various powers of $\left( {\textstyle \frac{1}{4}}\, g^4  ({ T^*}{}^{+})^2(T^-)^2 \right)^i$ available as deformations form a vector space in duality-satisfying  ${\cal N}=0$ theories.   Any value of the coefficients $a_{i}$ represents a valid duality satisfying theory.  But these vectors $\vec{a}$, up to trivial normalizations, completely distinguish theories which cannot be related by local field transformations.   From this point of view the fact that Bossard-Nicolai model has vanishing $a_i$ for all $i>0$ completely distinguishes it from the BI theory.
 
The models with manifest ${\cal N}=1$ supersymmetry and no space-time derivatives follow the same pattern as discussed in ref.~\cite{CKR}.


To discuss ${\cal N}=2$ we introduce the  notation of ref.~\cite{Kuzenko:2000uh,CKR,Neq2duality}. Many new examples of manifestly $\mathcal{N}=2$ supersymmetric models with self-duality were given in \cite{Neq2duality}. Some corresponding choices of the sources of deformation were given in the form
\begin{multline}\label{I}
{\cal I}(T^-, {\overline T}\,{}^{+})=\int {\rm d}^{12} \cZ \Bigg(
\lambda\; a_0 \; (T^-)^2 ({\overline T}\,{}^{+})^2\\
+  \lambda^2\; a_1\; (T^-)^3 \Box ({\overline T}\,{}^{+})^3
 + \lambda^3\;  a_2 (T^-)^4 \Box^2 ({\overline T}\,{}^{+})^4\\
+  \lambda^3 \; a_3 \; (T^-)^2 ({\overline T}\,{}^{+})^2 \SDbar((T^-)^2) \SD(  ({\overline T}\,{}^{+})^2) \\
+{\cal O}(\lambda^4)\Bigg)\,
\end{multline}
with $\cZ^A = (x^a, \q^\a_i, {\bar \q}^i_\ad) $, with $a$ and $\a$ being a vector and Weyl spinor Lorentz 
indices and $i = {1}, {2}$  being the $SU(2)$ R-symmetry  index.  Analogous combinations
\be
\label{def}
T^{+}= \cW- {\rm i} \; {\cal M}\, ,  \qquad  {\overline T}\,{}^{+} = \cW +  {\rm i} \; {\cal M}
\ee
\be
\label{def2}
T^{-}= \cWbar- {\rm i} \; \cMbar\, ,  \qquad {\overline T}\,{}^{-}= \cWbar  + {\rm i} \; \cMbar \ ,
\ee 
are given in terms of duality doublets, the superfields $\cW$ and ${\cal M}$, where ${\cal M}$ are treated independent of $\cW$.  Carrying out the deformation procedure of \cite{Neq2duality}
we recover the set of $\mathcal{N}=2$ self-dual actions depending on arbitrary parameters $\vec{a}$.  The Born-Infeld (and as we will argue hidden-supersymmetry) action from the literature requires a special choice of these parameters to reproduce the 
\bea
\vec{a}_{0-3}=(-2^{-4}, -2^{-6} \, 3^{-2}\, , - 2^{-12} \, 3^{-2}\, , 2^{10})
 \label{coefVector}
\eea	
Again, the $\mathcal{N}=2$ self-duality is valid for the reconstructive action defined by the sources of deformation in \eqn{I} for any choice of $\vec{a}$.  It should be noted that many more sources of deformation are available than those needed for the Born-Infeld ${\cal N}=2$ action -- see section IV.F of ref.~\cite{Neq2duality}.  Yet still the various independent deformation sources contributing to the ${\cal N}=2$ BI action represent a vector space distinguishing ${\cal N}=2$ duality satisfying theories unrelated by local field redefinitions for which we will argue only one vector represents hidden supersymmetry.



The $\mathcal{N}=2$ manifestly self-dual action proposed by Kuzenko-Theisen in \cite{Kuzenko:2000uh} has the following features. The  $U(1)$ $\mathcal{N}=2$ manifest self-duality condition is one of the requirements for constructing the model. The second condition is the  symmetry of the action under a certain non-linear symmetry with the bosonic parameter $\sigma$ under which the chiral $\mathcal{N}=2$ superfield ${\cal W}$ transforms as
\be
\delta {\cal W} = \sigma + \sigma \bar {\cal D}^4 \Bar {\cal Y}+ \bar \sigma \Box {\cal Y}
\ee
This includes the shift of a superfield by a constant parameter $\sigma$ and other ${\cal Y}$-dependent terms where  the chiral superfield ${\cal Y}$  is some unknown, in general,   functional of $\mathcal{N}=2$ superfields ${\cal W}$, $\bar {\cal W}$. The model in \cite{Kuzenko:2000uh} was constructed up to the level ${\cal W}^{8}$. To go beyond the ${\cal W}^{8}$ level in the action required establishing a higher level dependence of 
${\cal Y}$ on ${\cal W}$, $\bar {\cal W}$.  It was not clear from the construction if the model possessed any hidden supersymmetries.


A related development of the $\mathcal{N}=2$ manifestly supersymmetric action was suggested by Belluci, Ivanov, and Krivonos (BIK) in ref. \cite{Bellucci:2001hd} where the action was derived from the requirement that there is a spontaneously broken extra hidden $\mathcal{N}=2$ supersymmetry. The non-linear hidden $\mathcal{N}=2$ supersymmetry corresponds to the following transformations of the chiral superfield ${\cal W}$
\be
\delta {\cal W} = f (1-  {\frac{1}{2}} \bar { D}^4 \Bar {\cal A}_0)+ {\frac{1}{2}} \bar f \Box {\cal A}_0 + {\frac{1}{4 i}} \bar D^{i\dot \alpha} \bar f D_i^\alpha \partial _{\alpha\dot \alpha} {\cal A}_0
\ee 
with
\be
f= c+2 i \eta^{i\alpha} \theta_{i\alpha}\, , \qquad f= \bar c-2 i \eta^{i}_{\dot \alpha} \bar \theta^{\dot \alpha}_i
\label{HiddednSusyWTrans}
\ee
where the fermionic parameters of the spontaneously broken supersymmetry are $\eta^{i\alpha}, \bar \eta^i_{\dot \alpha}$ and the bosonic ones of the central charge are $(c, \bar c)$. The dependence of ${\cal A}_0$ on ${\cal W} $ requires the introduction of a chain of the superfields ${\cal A}_i$ which are identified as explicit recursive functions of the original superfields ${\cal W}$.  In this way the condition of an extra hidden $\mathcal{N}=2$ supersymmetry is supported recursively. However, in \cite{Bellucci:2001hd}  the condition of the $\mathcal{N}=2$ supersymmetric self-duality was not imposed and there is no obvious reason why it should  be present.

One of the reasons why all these actions with $\mathcal{N}=2$ and hidden  $\mathcal{N}=2$ supersymmetry were not completed into a closed form may be due to the manifest linearized $\mathcal{N}=2$ supersymmetry represented by the unconstrained superfields. As we will explain shortly, the complete action with $\mathcal{N}=2$ and hidden $\mathcal{N}=2$ supersymmetry \cite{BCKSV} has a deformed $\mathcal{N}=2$ supersymmetry. Therefore using undeformed and unconstrained $\mathcal{N}=2$ superfields may complicate the derivation of the action, rather than making it simple. This is a rather unexpected feature: in most situations  superfields  simplify the solutions of various problems. However, as discussed in ref.~\cite{BCKSV}, for the non-linear supersymmetries both the original Maxwell 
supersymmetry as well as the hidden one may have to undergo a deformation. This non-linear deformation violates the $\mathcal{N}=2$ Maxwell superfield structure and therefore the manifest undeformed $\mathcal{N}=2$ supersymmetry of the models in \cite{Kuzenko:2000uh,Bellucci:2001hd, Bellucci:2012nz} appears to be an obstacle  in attempts to  find a complete $\mathcal{N}=2$ supersymmetric Born-Infeld model.


A different approach, proposed in \cite{BCKSV}, is based on considering local $\kappa$-symmetry of superbrane actions acting on models with $\mathcal{N}=2$ and hidden $\mathcal{N}=2$ supersymmetry The relevant superbrane action, with a local $\kappa$-symmetry  
is in the class of {\it vector branes} models~\cite{BCKSV}; their world-volume dynamics is described by a vector multiplet. 
In  our case, regarding four dimensional $\mathcal{N}=2$ + $\mathcal{N}=2$ supersymmetry, it would be a V3 superbrane model -- an extended object associated with the  dimensional reduction of the V5 space-filling brane of a chiral IIb $(2,0)$ $D=6$ supergravity  \cite{Bergshoeff:2012jb}.  The complete action is presented in \cite{BCKSV}, which means that it can also be given as an expansion to all orders of deformation.  

We propose a conjecture here that  hidden supersymmetry  implies duality. We will argue that this  hidden supersymmetry happens if and only if the bosonic action is a Born-Infeld one and the  fermionic one is a Volkov-Akulov one (or related by local non-linear field redefinitions).

In our study of the relation between hidden supersymmetry and duality we take into account the following. In a model with a given amount of non-linear supersymmetries acting on the same set of fields, for example $\mathcal{N}=2$ + $\mathcal{N}=2$ models 
\cite{Kuzenko:2000uh,Bellucci:2001hd,Neq2duality}  one should take into account that manifest $\mathcal{N}=2$ supersymmetry with $\mathcal{N}=2$ superfields involves an auxiliary field. When the auxiliary field from the $\mathcal{N}=2$ superfield ${\cal W}$  is excluded on its equation of motion, the model depends on physical fields of an $\mathcal{N}=2$ Maxwell multiplet: a vector, a spinor and a complex scalar. Models in \cite{Kuzenko:2000uh,Bellucci:2001hd} coincide up to ${\cal W}^8$, whereas in \cite{Neq2duality} many more models with manifest $\mathcal{N}=2$ self-duality are presented. Only some of them coincide with the BIK action and therefore have the second hidden $\mathcal{N}=2$ supersymmetry.

Our conjecture on hidden supersymmetry and duality is based on the following observations.

{\it Observation 1}.  Even though there are a large number of distinguishable  self-dual actions with manifest  ${\cal N}=1$ supersymmetry,  only one (up to local\footnote{Only models sharing symmetries may be related to  each other via a non-linear local change of field variables, e.g. $A_\mu' = A_\mu' (A_\mu)$.   One can craft relations between distinct  solutions of the self-duality condition but this would involve transformations {\it non-local} in the field variables, i.e. transformations taking the form $t'=t'(t,z)$ and $ z'= z'(t, z)$ with  $t= {\frac{1}{4}} {F^2}$, $z= {\frac{1}{4}} F\tilde F\, $.} change of variables and trivial normalization) has the superfield action with an extra hidden  ${\cal N}=1$ supersymmery.  We claim the directly analogous structure holds for ${\cal N}=2$.

{\it Observation 2}.  The  $\mathcal{N}=2$ + $\mathcal{N}=2$  V3 model~\cite{BCKSV} has very simple properties:  complete to all orders, when truncated to pure vectors, it is a BI model 
\begin{equation}
S^{V3}|_{\phi=\lambda=0}=  \frac{1}{\alpha^2} \int d^{4} x\,\left\{1-
\sqrt{- \det (\eta_{\mu\nu} + \alpha  F_{\mu\nu})}  \right\} \, .
\label{V3BI}\end{equation}
I.e. when the fermions $\lambda$ and the scalars $\phi^I$ are absent it becomes a BI action with clear self-duality property
$F\tilde F+G\tilde G=0$.

When we truncate the 2-forms and the scalars covariantly from the V3 action we find that
it depends on fermions as a Volkov-Akulov type action for a Goldstino
\begin{equation}
S^{V3}|_{{\cal F}_{\mu\nu}=\Pi_\mu^I=0} = \frac{1}{\alpha^2} \int d^{4} x\,\left\{1
- \det  (\delta^{\mu}_{\nu}
-\alpha^2\bar\lambda \Gamma^{\mu}
\partial_\nu \lambda)  \right\}\ . 
\label{V3VA}\end{equation}
According to \cite{Kuzenko:2005wh} this then satisfies the fermionic part of the  ${\cal N}=1$ self-duality condition with account of a non-linear Goldstino field change of the variables.   We suggest that this model is related by local field redefinitions to the BIK action.

Our conjecture `hidden supersymmetry implies duality'  explains  why the BIK action up to ${\cal W}^8$  in \cite{Bellucci:2001hd} with established hidden second supersymmetry coincides with the action in \cite{Kuzenko:2000uh} with established manifest self-duality and predicts that  the  ${\cal W}^{10}$ terms in the BIK action  has a manifest $\mathcal{N}=2$ self-duality, which we now verify.   
Happily a much more elegant option than ``brute-force'' calculation is available to us.  We simply find the duality-consistent sources of deformation necessary to generate the action through order ${\cal W}^{10}$.  We find that only two additional  sources of order $\lambda^4$ are required: 
\begin{multline}
{\cal I}_{\lambda^4}(T^-, {\overline T}\,{}^{+})= \lambda^4  \int {\rm d}^{12} \cZ \Bigg(
\; a_4 \; (T^-)^5 \Box^3 ({\overline T}\,{}^{+})^5\\
+ 
\; a_5  \;    (T^-)^3 \SD( ({\overline T}\,{}^{+})^2) \Box ({\overline T}\,{}^{+})^3 \SDbar((T^-)^2) \,\Bigg).
\end{multline}
The  values of the new coefficients required to reconstruct the BIK action through known orders: 
\bea
\vec{a}_{4,5}=(  2^{-5}\, 3^{ -2} \, 5^{ -2} ,\, 2^{ -3}\, 3^{-2} )\,.
 \label{coefVector2}
\eea
These coefficients could be any numbers and still satisfy duality, but these and those given in \eqn{coefVector} are the particular values required to reconstruct the BIK action through order ${\cal W}^{10}$, and thus we argue possess a particular hidden-supersymmetry.  The discovery of these deformation sources is all that is required to demonstrate that the BIK action, through this order, is a solution to the self-duality condition.  Of course one can also directly verify this property of the action by checking the self-duality constraint  which we elide to an auxiliary file~\cite{auxFile}.    


In conclusion, in models of $\mathcal{N}=2$ supersymmetry  with a hidden second $\mathcal{N}=2$ supersymmetry, self-duality is a feature, whereas generic manifest $\mathcal{N}=2$ supersymmetric and self-dual models do not necessarily come with a second hidden supersymmetry.   We showed constructively that the ${\cal W}^{10}$ action  of  \cite{Bellucci:2001hd} is self-dual. 
The dramatic cancellations~\cite{auxFile} between the many various terms in this action's explicit satisfaction of the self-duality condition is further explained by the  existence of the complete  model with $\mathcal{N}=2$ + hidden $\mathcal{N}=2$ supersymmetries of ref.~\cite{BCKSV}.  It will be interesting to see if this model can be entirely understood to all orders at the level of duality preserving sources of deformation as was done for the purely bosonic BI case~\cite{Aschieri:2013nda}.

\vskip .2 cm 

We thank  E.~Bergshoeff, J.~Broedel, W.~Chemissany, F.~Coomans, S.~Ferrara, T.~Ortin, C.~S.~Shahbazi, R.~Roiban, and A.~Van Proyen for related collaboration.  We are especially grateful to J.~Broedel, S. Ferrara, and R.~Roiban for comments on earlier drafts.  We would also like to thank P.~Aschieri, G.~Gibbons, Z.~Komargodski and A.~Tseytlin for interesting conversations.  This research was supported by the US National Science Foundation under  PHY-0756174,  the Stanford Institute for Theoretical Physics, and a grant from the John Templeton Foundation.

\end{document}